\documentclass[sigconf, authoryear]{acmart}

\AtBeginDocument{%
}

\copyrightyear{2025}
\acmYear{2025}
\setcopyright{rightsretained}
\acmConference[ASSETS '25]{The 27th International ACM SIGACCESS Conference on Computers and Accessibility}{October 26--29, 2025}{Denver, CO, USA}
\acmBooktitle{The 27th International ACM SIGACCESS Conference on Computers and Accessibility (ASSETS '25), October 26--29, 2025, Denver, CO, USA}
\acmDOI{10.1145/3663547.3759738}
\acmISBN{979-8-4007-0676-9/2025/10}

\usepackage{graphicx}
\usepackage{float}
\usepackage{subfigure}
\usepackage{subcaption}
\usepackage{hyperref}

\begin{document}

\title{\textsc{RunPacer}: A Smartwatch-Based Vibrotactile Feedback System for Symmetric Co-Running by Visually Impaired Individuals and Guides}

\author{Yichen Yu}
\authornote{Both authors contributed equally to this research.}
\orcid{0009-0001-0175-3253}

\affiliation{%
  \institution{AYXR Research Group}
  \institution{North Carolina State University}
  \city{Raleigh}
  \state{North Carolina}
  \country{USA}
}
\affiliation{%
  \institution{Georgia Institute of Technology}
  \city{Atlanta}
  \state{Georgia}
  \country{USA}
}
\email{lunarsboy@gmail.com}

\author{Huan-Song Xu}
\authornotemark[1]
\orcid{0009-0004-6488-114X}

\affiliation{%
  \institution{Department of Computer Science and Information Engineering}
  \institution{National Taiwan University of Science and Technology}
  \city{Taipei}
  \country{Taiwan}
}
\email{m11215036@mail.ntust.edu.tw}

\author{Ming-Yen Lin}
\orcid{0000-0003-3180-3132}

\affiliation{%
  \institution{Department of Information Engineering and Computer Science}
  \institution{Feng Chia University}
  \city{Taichung}
  \country{Taiwan}
}
\email{linmy@fcu.edu.tw}

\renewcommand{\shortauthors}{Yu, Xu and Lin}

\begin{abstract}
Visually impaired individuals often require a guide runner to safely participate in outdoor running. However, maintaining synchronized pacing with verbal cues or tethers can be mentally taxing and physically restrictive. Existing solutions primarily focus on navigation or obstacle avoidance but overlook the importance of real-time interpersonal rhythm coordination during running. We introduce \textsc{RunPacer}, a smartwatch-based vibrotactile feedback system that delivers synchronized rhythmic pulses to both runners. In contrast to conventional guide-running systems that rely heavily on continuous verbal communication or mechanical tethering, \textsc{RunPacer} emphasizes interpersonal cadence alignment as its core interaction model. By pre-setting a target step frequency or dynamically adapting to the guide's natural pace, the system ensures that both runners receive identical haptic cues, enabling them to maintain coordinated motion intuitively and efficiently. This poster presents the system architecture, positions it within prior research on haptic entrainment, and outlines the vision for future field deployment, including potential multimodal feedback extensions. \textsc{RunPacer} contributes a lightweight, socially cooperative, and non-visual assistive framework that reimagines co-running as a shared, embodied, and accessible experience.
\end{abstract}

\ccsdesc[500]{Human-centered computing~Accessibility technologies}
\ccsdesc[500]{Human-centered computing~Ubiquitous and mobile computing systems and tools}
\ccsdesc[300]{Hardware~Haptic devices}

\keywords{Accessibility, Assistive Technology, Vibrotactile Feedback, Wearable Computing, Haptic Interaction, Co-running, Smartwatch}

\maketitle

\section{Introduction}
Running is a universally accessible form of physical activity that provides a multitude of physiological and psychological benefits, such as enhancing cardiovascular health, reducing anxiety and stress, improving sleep quality, and supporting emotional well-being. For many, running is not only a form of exercise but also a means of self-expression and social connection. However, for individuals with visual impairments, the experience of running outdoors offers distinct and persistent challenges. These challenges are not merely physical but are intricately connected with issues of autonomy, communication, and safety.

In typical scenarios, visually impaired runners must rely on a human guide, often tied by a short rope or connected by a series of verbal cues. Synchronization between the guide and the visually impaired runner is essential but often difficult to maintain. Verbal guidance demands constant attention and clear environmental acoustics, which may not always be feasible in outdoor environments. Moreover, the use of physical tethers can restrict natural stride, limit arm swing, and introduce discomfort, ultimately disrupting the running rhythm and reducing user agency. These traditional methods also reinforce a unidirectional dependency model, where the guide dictates the pace and direction, leaving the visually impaired individual with limited control over their experience \cite{xu2017configurable, lind2020wearable}.

Emerging research in motor coordination and haptic feedback systems has demonstrated potential in reducing these barriers. Studies have demonstrated that rhythmic tactile cues, when delivered with precision, can reinforce movement regularity and support spontaneous synchronization between individuals. This phenomenon, known as entrainment, reflects the human body’s innate ability to synchronize with external rhythmic stimuli and has been observed across a range of social and locomotor activities \cite{islam2022vibrotactile, schroeder2025amplitude, georgiou2020rhythmic}.

To leverage this potential, we present \textsc{RunPacer}, an innovative smartwatch-based system intended to promote real-time, bidirectional synchronization between a guide and a visually impaired runner. Rather than relying on verbal instructions or physical tethers, \textsc{RunPacer} employs shared vibrotactile feedback, delivering synchronized step-based pulses to both participants. This approach offers a seamless, intuitive, and cognitively light alternative to traditional methods, empowering visually impaired enabling users to participate in running experiences that are not only safer and more comfortable but also more autonomous, social, and enjoyable \cite{baldi2020wearable, shull2015haptic, ballardini2020vibrotactile}.

\section{Related Work}
Researchers have long explored the use of haptic and vibrotactile feedback to guide motion in rehabilitation, exercise, and assistive technologies, aiming to support users with varying motor or sensory limitations. A considerable body of work has demonstrated that haptic cues can be pivotal in motor learning, spatial orientation, and rhythm synchronization. For example, Xu et al. developed a wearable vibrotactile system that improves postural control through directional feedback, showing encouraging outcomes in both static balance tasks and dynamic movements \cite{xu2017configurable}. Similarly, Kalff et al. found that gait-synchronized vibrotactile feedback enhances mobility results in subjects with lower-limb amputations, demonstrating how real-time tactile feedback can facilitate neuromotor rehabilitation and entrainment \cite{kalff2024impact}.

Beyond medical rehabilitation, haptic feedback has been increasingly integrated into locomotion assistance systems. In the domain of gait training, De Angelis et al. performed an exhaustive analysis and concluded that vibrotactile stimulation not only supports postural stability but also encourages proactive participation by providing non-intrusive cues that users can internalize \cite{de2021vibrotactile}. Mahmud et al. extended this notion to immersive environments by incorporating vibrotactile signals into virtual reality standing balance training programs, showing that tactile stimuli could effectively substitute or augment visual cues \cite{mahmud2022standing, mahmud2022vibrotactile}.

Wearable haptic systems have also been used to facilitate socially coordinated behaviors. Lisini et al., for instance, proposed a wearable solution for remote social walking, allowing pairs of persons to preserve synchronized step rhythms even while being geographically separated \cite{baldi2020wearable}. Their work illustrates the potential of vibrotactile entrainment in interpersonal contexts, suggesting broader applications beyond solo rehabilitation, including shared physical activities like dancing, walking, and running.

However, most prior work focuses on navigation, balance, or motor control in constrained indoor settings, often with the goal of compensating for deficits or preventing falls. Assistive devices designed for the visually impaired, such as Wayband and other commercial systems, have largely targeted directional guidance, providing users with cues for pathfinding or obstacle avoidance through vibrotactile feedback embedded in wearables like shoes or belts. While effective for spatial navigation, these systems are not optimized for interpersonal movement synchronization, particularly in high-tempo and coordination-intensive scenarios like running with a guide \cite{lind2020wearable, islam2018wearable}.

Moreover, despite the exhibited promise of entrainment theory in social and sensorimotor contexts, few systems have explicitly operationalized this concept into shared, bidirectional pacing tools for co-locomotion. The majority of existing solutions emphasize asymmetrical interaction models, where one user leads and the other passively follows. This dynamic limits autonomy and co-agency, especially for visually impaired individuals who may wish to take a more active role in the running experience \cite{islam2022vibrotactile}.

Our work bridges multiple strands of this literature—rehabilitation, wearable haptics, social entrainment, and assistive technology—to fill a void: the lack of tools that support symmetric, co-tempo running experiences for visually impaired individuals. \textsc{RunPacer} introduces a new category of assistive systems that use synchronized haptic feedback not for navigation, but for enabling mutual rhythm alignment. This repositioning shifts the emphasis from guidance to collaboration, reframing the role of the assistive device as a facilitator of shared agency and embodied communication \cite{shull2015haptic, ballardini2020vibrotactile}.

\section{System Design}

\begin{figure}[H]
  \centering
  \includegraphics[width=\linewidth]{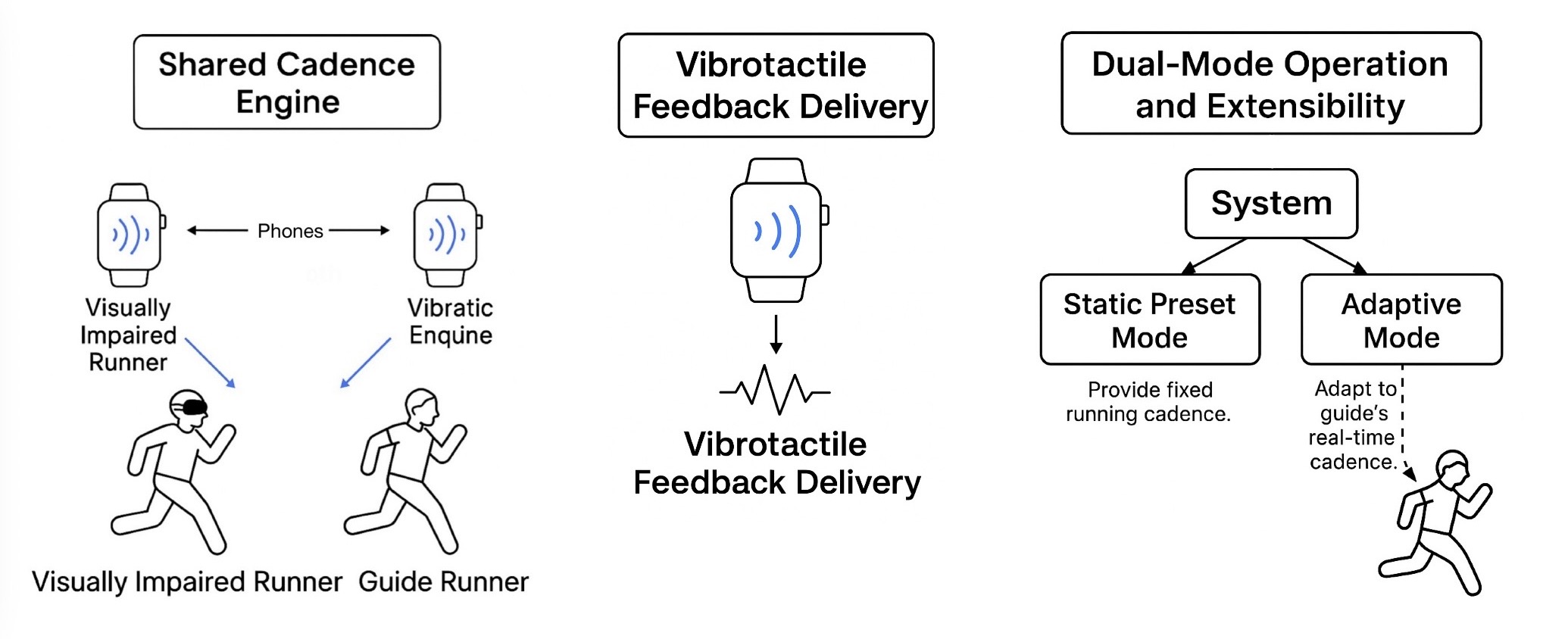}
  \caption{Overview of the \textsc{RunPacer} system, composed of three interconnected modules that operate in tandem to support synchronized co-running.}
  \label{fig1}
\end{figure}

\textsc{RunPacer} is built around three primary modules(see Figure~\ref{fig1}) designed to provide precise, intuitive, and low-latency feedback that supports co-running between visually impaired individuals and their guides. The system emphasizes temporal alignment, minimal cognitive intrusion, and cross-device compatibility to ensure robust usability in diverse real-world contexts.

\subsection*{Shared Cadence Engine}
Shared Cadence Engine, is responsible for generating and broadcasting a synchronized step rhythm to both users' smartwatches. This engine supports two operational modes: a manually configured cadence (e.g., 160 steps per minute) for consistent training environments, and a dynamic mode that continuously detects the guide's real-time step frequency through onboard inertial sensors.

In either case, a synchronized pulse stream is transmitted to both runners via a low-latency Bluetooth channel. The synchronization system is engineered to operate with under 100 milliseconds of delay, which is essential for preserving the perceptual integrity of the rhythm and maintaining entrainment \cite{schroeder2025amplitude}. The cadence engine is modular and platform-agnostic, supporting integration across smartwatch ecosystems using standardized communication APIs.

\subsection*{Vibrotactile Feedback Delivery}
Vibrotactile Feedback Delivery, transforms the rhythmic signals into distinct, synchronized pulses using the built-in haptic motors of each smartwatch \cite{georgiou2020rhythmic}. These pulses are short, crisp, and calibrated to provide perceptual clarity without becoming intrusive or fatiguing.

The key design parameters include pulse duration, intensity, and waveform configuration. The pulse duration is optimized between 100–200 milliseconds to ensure clarity during motion. Pulse intensity is adjustable to accommodate ambient noise levels, individual preferences, and individual skin sensitivity \cite{xu2017configurable}. The pulse waveform is configured to deliver single-tap pulses, which were favored over longer or multi-tap patterns in our preliminary user testing.

These carefully tuned parameters enable the vibrotactile signals to act as an intuitive, shared tempo for both runners, supporting precise rhythm alignment without the need for active interpretation.

\subsection*{Dual-Mode Operation and Extensibility}
Dual-Mode Operation and Extensibility, adds robustness for diverse running scenarios. In Static Preset Mode, the system provides a fixed cadence suitable for controlled training or rehabilitation sessions. In Adaptive Mode, it continuously monitors the guide’s running cadence and mirrors it for the visually impaired runner in real-time, promoting seamless co-adaptation.

By adopting a modular architecture, \textsc{RunPacer} is designed not only to meet current assistive needs but also to serve as a foundation for future innovations in wearable-based, co-locomotion support systems.

\section{Implementation}

We developed the initial implementation of \textsc{RunPacer} using Apple Watch Series 7 with a custom application programmed in Swift. The app leverages Core Motion and Core Bluetooth frameworks for step detection and real-time communication. Cadence tracking accuracy was enhanced through smoothing algorithms and predictive filtering techniques, resulting in stable synchronization signals with a measured latency of under 100 milliseconds during testing. The smartwatch’s Taptic Engine provided precise vibration pulses configured via custom waveform patterns to optimize perceptibility and minimize habituation.

The implementation was designed to be flexible and extensible, enabling future integration of additional physiological sensors such as heart rate monitors, electromyography armbands, or fatigue-detection modules. This adaptability aims to facilitate personalized and context-aware pacing strategies, further enhancing \textsc{RunPacer}’s utility for visually impaired runners across varied running environments and training regimens.

\section{Heuristic Evaluation}

We conducted a preliminary heuristic evaluation to assess \textsc{RunPacer}’s usability and effectiveness. Ten participants formed five pairs, each consisting of one blindfolded runner and one guide. Participants ran together on a 200-meter straight track using only \textsc{RunPacer}’s vibrotactile feedback for synchronization.

Post-run evaluations included Likert-scale questionnaires assessing synchronization ease, clarity of vibration cues, comfort, perceived safety, and cognitive load compared to traditional verbal methods. Qualitative interviews explored user perceptions of autonomy, ease of use, and overall satisfaction. Participants overwhelmingly reported that the system significantly reduced cognitive load and enhanced their sense of autonomy and safety compared to verbal cues or tethering.

However, some participants suggested improvements such as adjustable vibration intensity for different running speeds, and integration of subtle auditory feedback as an additional optional modality to support rhythm synchronization during high-speed or complex running scenarios.

\section{Future Work}

Our next steps involve expanding \textsc{RunPacer}’s adaptability and validating its effectiveness through real-world deployment. We plan to conduct field studies with visually impaired runners and guide partners across diverse environments to assess synchronization accuracy, usability, and long-term comfort.

We also aim to integrate physiological sensing such as heart rate or electromyography to support fatigue-aware pacing adjustments. This would allow \textsc{RunPacer} to dynamically adapt vibration patterns based on exertion levels, further personalizing the co-running experience.

Additionally, we are exploring multimodal feedback by combining bone-conduction audio with vibrotactile cues to reinforce rhythm without increasing cognitive load. Another direction is scaling the system for group running, enabling synchronization among multiple runners through shared feedback streams.

Ultimately, we envision \textsc{RunPacer} as a customizable platform. We plan to release a developer toolkit to support broader use in therapy, fitness, or collaborative training contexts.

\section{Conclusion}

\textsc{RunPacer} was developed specifically to address the limitations of traditional guide-running methods such as verbal cues and physical tethering, which can limit autonomy, impair comfort, and impose cognitive burden on visually impaired runners. 

Our initial prototype, implemented on commercial smartwatch hardware, demonstrated reliable performance in lab-based and preliminary field scenarios. Participants reported increased autonomy, reduced cognitive load, and greater comfort in contrast to conventional verbal methods.

The contribution of \textsc{RunPacer} lies not only in its specific implementation, but also in the broader design implications it offers for inclusive movement interfaces. It reframes co-running as a bidirectional, rhythmic, and social interaction, leveraging entrainment principles and vibrotactile cues to support mutual adaptation rather than passive following. This reframing has possible implementations beyond running, including co-walking, synchronized rehabilitation, or embodied collaborative learning \cite{ballardini2020vibrotactile}.

Looking forward, we envision \textsc{RunPacer} as a flexible platform for inclusive embodied computing, adaptable to diverse contexts, populations, and movement goals. Through continued development, field validation, and open-source dissemination, we aim to foster a more equitable and expressive future for physical activity, where movement becomes not just accessible, but collaboratively shared.

\bibliographystyle{ACM-Reference-Format}
\bibliography{reference}

\end{document}